\documentclass[fleqn,12pt,twoside]{article}
\usepackage{espcrc1}

\usepackage{graphicx}
\usepackage[figuresright]{rotating}

\newcommand{\AmS}{{\protect\the\textfont2
  A\kern-.1667em\lower.5ex\hbox{M}\kern-.125emS}}
\newcommand{\be}{\begin{equation}}
\newcommand{\ee}{\end{equation}}
\newcommand{\bea}{\begin{eqnarray}}
\newcommand{\eea}{\end{eqnarray}}
\newcommand{\gton}{\mathrel{\lower.9ex
                  \hbox{$\stackrel{\displaystyle >}{\sim}$}}}
\newcommand{\lton}{\mathrel{\lower.9ex
                  \hbox{$\stackrel{\displaystyle <}{\sim}$}}}

\title{The deconfined phase near $T_c$ and its decay into hadrons}

\author{A.~Dumitru\address[MCSD]{Physics Department,
Brookhaven National Lab, Upton, NY 11973, USA}%
        \thanks{Supported by the DOE Grant DE-AC02-98CH10886.
e-mail: dumitru@quark.phy.bnl.gov},
        R.~D.~Pisarski\addressmark}
       
\begin{document}

\maketitle

\begin{abstract}
We sketch an effective theory for the deconfined state of QCD near
$T_c$. This relates the behavior of the expectation value of the
Polyakov loop, and its two-point functions, to the pressure.
Defining the ``mass'' of three and two gluon states from the imaginary and real
parts of the Polyakov loop, while this ratio is 3:2 in perturbation
theory, at $T_c^+$ it is 3:1.
We also discuss the decay of the deconfined state into hadrons.
\end{abstract}

\section{Introduction}

High-energy inelastic processes provide the unique opportunity to 
recreate the high energy density state of QCD matter that presumably
prevailed during the first microseconds of the evolution of the early
universe: the so-called Quark-Gluon Plasma (QGP).
Simply speaking, the QGP is the state of matter for which quarks and gluons are
deconfined and chiral symmetry is restored.
Even if produced, the QGP is a transient state of matter which can not be
observed directly. Experimentally, one observes hadrons which are produced from
the decay of the deconfined QGP state.
The deconfined state at asymptotic temperatures can presumably be
understood in terms of a weakly interacting gas of quasiparticles. It is
rather crucial, however, to develop some understanding of QCD {\em near}
the deconfinement temperature $T_c$ in order to be able to interpret its
``fingerprints''.

\section{Polyakov Loop Model}
The basic idea behind the ``Polyakov loop model'' is that the
free energy of the deconfined state of $SU(3)$ gauge theory
is dominated by a condensate of gauge invariant Polyakov
loops~\cite{Pisarski:2000eq}
\be\label{PL_loop}
\ell(\vec{x}) \equiv \frac{1}{3} \; {\rm tr}\; L(\vec{x}) =
 \frac{1}{3} \; {\rm tr}\; {\cal P} 
\exp \left( i g \int^{1/T}_0 A_0(\vec{x},\tau) \, 
d\tau \right)~.
\ee
It is based on the observation from the lattice
that the Polyakov loop becomes light near $T_c$~\cite{pol_mass}.
Here, $A_0$ is in the fundamental representation and so can not acquire
an expectation value since that would break the gauge symmetry.
In contrast, the Wilson line $L$
transforms as an adjoint field under the local $SU(3)/Z(3)$ 
gauge symmetry, and as a field with charge one under the global
$Z(3)$ symmetry. Thus, it can acquire an expectation value within the center
of the group and so represents an order parameter for deconfinement:
$\langle\ell\rangle=0$ corresponds to the confined phase in which
the t'Hooft $Z(3)$ symmetry is restored, while $\langle\ell\rangle\ne0$
above $T_c$.

Near $T_c$, instead of dealing with $A_0$,
we construct an effective Lagrangian directly for the
Polyakov loop itself.
The theory must be invariant under the global $Z(3)$ symmetry.
For the present purposes, we
take the effective Lagrangian to be
\be \label{eff_L}
{\cal L} = 
|\partial_\mu \ell|^2 
- \left(- \frac{b_2(T)}{2} |\ell|^2 - \frac{b_3}{3}
\left( \frac{\ell^3 + (\ell^*)^3}{2} \right)
+ \frac{1}{4}\left( |\ell|^2\right)^2\right)b_4 T^4~.
\ee
(Here, we have suppressed all 
renormalization constants for the wavefunction, mass and coupling constants.)
The terms which only involve powers of $|\ell|^2$ are
each invariant under a global symmetry of 
$U(1)$, and so is the kinetic term.
The cubic term, $\sim b_3 \ell^3$, is only 
invariant under the global $Z(3)$ symmetry.  
We perform a mean field analysis, where
all coupling constants are taken as constant with temperature, except
for the mass term, $\sim b_2 |\ell|^2$.  
About the transition, condensation
of $\ell$ is driven by changing the sign of $b_2$.
Also, $b_3>0$ so that in the $Z(3)$ model,
there is always one vacuum with a real, positive expectation value for
$\ell$. 

The couplings $b_2(T/T_c)$ and $b_3$ could, in principle, be computed on the
lattice ($b_4$ is fixed by the asymptotic number of degrees of freedom)
once the wavefunction renormalization for $\ell$ is known. Alternatively,
they can be deduced by matching some physical observable, for example the
pressure or the screening
mass for $\ell$ as obtained from~(\ref{eff_L}), to lattice
data; for details we refer to~\cite{adrob_ptfluc,ove_Kpi}.

Fermions in the fundamental representation break the global $Z(3)$
explicitly such that $\langle\ell\rangle\ne0$ below $T_c$. Assuming that
those fields can be integrated out, a term linear in $\ell$, $-b_1(\ell+
\ell^*)/2$, arises in the effective Lagrangian. 
The lattice finds that near $T_c$, the temperature-susceptibility for the
Polyakov loop peaks strongly~\cite{digal}, suggesting that $b_1$ is small.
Further, the pressure appears to be ``flavor independent'':
$p/p_{\rm id}$, where $p_{\rm id}$ is the ideal-gas pressure, is a
nearly {\em universal} function of $T/T_c$~\cite{Karsch:2000ps}.
This also indicates
that $b_1$ is small, and that $b_2(T/T_c)$ is a universal function. If so,
correlation lengths in QCD increase near $T_c$.

\section{Two-point functions of $SU(3)$ Polyakov Loops}
In perturbation theory, the Polyakov loop is near unity,
and can be expanded in powers of $A_0$. For example, if the
$A_0$ field is purely static,
\be
\ell \; \approx \; 1 \; - \; \frac{g^2}{3 T^2} \; {\rm tr}
\left( A_0^2 \right) \; - \; i \; \frac{g^3}{3 T^3} \;
{\rm tr} \left( A_0^3 \right) \; + \ldots 
\ee
Thus the real part of the Polyakov loop starts out as a coupling to 
two $A_0$'s, and the imaginary part of the Polyakov loop, to three
$A_0$'s.  Thus to lowest order in perturbation theory,
$\langle \ell_{r}(x) \ell_{r}(0) \rangle - \langle \ell \rangle^2 \sim 
{\exp(- 2 m_D x)}/{x^2}$, $\langle \ell_{i}(x) \ell_{i}(0) \rangle \sim 
{\exp(- 3 m_D x)}/{x^3}$,
with $m_D=gT$ the Debye mass. Thus, if one
defines $m_r$ and $m_i$ from the exponential falloff of the two-point
functions, we conclude that the
masses for the real and the imaginary parts of $\ell$ behave as
$m_i/m_r = 3/2$ (this ratio is independent of the wavefunction renormalization
of $\ell$~\cite{2point}). In turn, $m_{i,r}(T)$ can be obtained from the
curvature of the effective potential~(\ref{eff_L}), and thus can be related to
the behavior of either the expectation value $\langle\ell\rangle$ or the
pressure $p$. For example, for very small $b_3$ one has the simple
relation~\cite{adrob_eloss} $(m_r/T)^2\sim\langle\ell\rangle^2\sim\surd
p/p_{id}$, and so indeed the mass becomes small for $T\to T_c^+$. 
Moreover, for any $b_3>0$ and any function $b_2(T/T_c)$, at $T_c^+$
one finds $m_i/m_r=3$ (plus corrections from pentic and hexatic interactions
of $\ell$~\cite{2point}). This prediction deviates strongly from the
counting in perturbation theory, where three (exchanged) electric gluons are
3/2 times heavier than two gluons. From~(\ref{eff_L}) though, at $T_c^+$ three
gluons are {\em three} times heavier
than two gluons. Recent lattice studies~\cite{Datta:2002je} in $SU(3)$
gauge theory do find that the mass ratio of a three versus a two-gluon state
increases sharply from $\approx1.5$
at high temperature to $\approx 2.5$ near $T_c$.

\section{The decay of the Condensate for $\ell$ into hadrons}
The effective Lagrangian now allows to study how the condensate for $\ell$
decays into hadrons at $T_c$. The potential becomes rather flat
at $T_c$, as seen from the growth of the correlation length $\xi_\ell$.
On the other hand, $\xi_\ell$ decreases rapidly both below and above $T_c$.
Within the Polyakov loop model, this implies a rather small discontinuity
of $\langle\ell\rangle$ at $T_c$ followed by a rapid growth. Thus, the
{\em shape} of the effective potential must change very rapidly near
$T_c$~\cite{adrob_ptfluc}. Consequently, it is rather likely
that the long wavelength modes of $\ell$ do not stay in equilibrium
as the system cools
through $T_c$ (for actual real-time numerical solutions of the dynamics
see~\cite{ove_bub}). This leads to a rapid decay of the condensate for
$\ell$ into hadrons,
termed  ``explosive''~\cite{adrob_ptfluc,ove_bub} for the fact that
the transition occurs in an ``instant'' throughout the
entire volume (i.e., on a space-like surface). In contrast, if equilibrium
were maintained during the transition,
\underline{generic local conservation laws} for
energy and momentum demand that the deconfined phase be ``consumed''
by a propagating burning front (on a time-like surface) with a small velocity
for either a first-order transition or a rapid crossover;
this behavior was argued to be incompatible with
RHIC HBT results~\cite{gyulassy}. The only {\em pre}diction of
$R_{out}\simeq R_{side}$ is from a sudden time-like transition~\cite{CC},
which can be based on the above-mentioned behavior of the effective
potential for the Polyakov loop~\cite{adrob_ptfluc,ove_bub}.

Another interesting issue, besides real-time dynamics, is the chemical
composition of the confined state, i.e.\ the relative yields of various
hadron species. It is rather remarkable that the hadronic states appear to
be populated statistically according to the available phase
space~\cite{statis}. This could be due to rapid equilibration
among the hadron species after hadronization. One would then naturally
expect that the temperature for freeze-out is significantly lower than that
for hadronization.
However, thermodynamic fits to the measured hadron
multiplicities at RHIC yield freeze-out temperatures essentially {\em equal}
to the
deconfinement temperature $T_c$. That is, the chemical composition
is determined at hadronization, with little time for
equilibration processes. This is confirmed by detailed studies of hadronic
rescattering after hadronization within kinetic theory~\cite{bass}.

In contrast,
in the Polyakov loop model, chemical composition is dynamically determined
by the decay of the condensate for $\ell$ into
hadrons~\cite{ove_Kpi}. Coupling $\ell$ to some chiral multiplets $\Phi$,
the occupation number of each state will be determined by the mass of that
state relative to the mass of $\ell$ at (or near) $T_c$, when the expectation
value $\langle\ell\rangle$ in the block-spin averaged effective theory
``rolls over'' from $\langle\ell\rangle\ne0$ to $\langle\ell\rangle=0$.
Clearly, if say $m_\ell<m_\pi$, this will result in a vast suppression of
all heavier states like the kaon relative to the pion. If, in turn,
$m_\ell>m_K$, the phase space for the decay into pions and kaons will
be similar, resulting in $K:\pi$ near unity.
To reproduce the experimental result of $K:\pi\approx1:10$, one must have
$m_\pi<m_\ell<m_K$. Indeed, this is the
case: fixing $b_2(T/T_c)$ below $T_c$ from lattice data on the string tension
in $SU(3)$ pure-gauge theory, and assuming
$b_1\approx0$, $m_\ell\approx400$~MeV just below $T_c$.
(In~\cite{ove_Kpi} the mass shifts from fluctuations
of the chiral fields where taken into account by means of a Hartree-type
factorization but do not change the above ordering). On the other hand,
assuming that the free energy near $T_c$ is dominated by
quasi-gluons with masses $500-1000$~MeV which then decay into $\pi$, $K$
etc.\ should lead to $K:\pi\approx1$~\cite{Biro:vj}.

In thermal QCD the mass of $\ell$
also has the meaning of an inverse correlation length $\xi_\ell = 1/m_\ell$. 
From the above, $\xi_\ell\simeq 1/2$~fm just below $T_c$, which is a
relatively large correlation length on hadronic scales.

The above arguments could be tested in $pp$ collisions at RHIC.
High-multiplicity $pp$ events can have energy and particle densities
comparable to those of central $Au+Au$ collisions, as measured by
$dN_{ch}/d\eta/A^{2/3}$. 
In $pp$ events, however, correlation lengths can not
become large because the system is small.
This will produce different relative hadron yields at hadronization.
Thus, it would be interesting to know how the $K:\pi$ ratio etc.\
in high-multiplicity $pp$ collisions compares to that in $Au+Au$.

\end{document}